%% file: acm_nanocom_main.tex
\documentclass[sigconf]{acmart}
\usepackage[printonlyused]{acronym}
\usepackage{subfigure}
\usepackage{hyperref}
\usepackage{bm}
\usepackage{fixltx2e}
\usepackage{mathtools}
\usepackage[font=small,labelfont=bf]{caption}
\DeclarePairedDelimiter\ceil{\lceil}{\rceil}

\renewcommand\footnotetextcopyrightpermission[1]{}

\def\BibTeX{{\rm B\kern-.05em{\sc i\kern-.025em b}\kern-.08emT\kern-.1667em\lower.7ex\hbox{E}\kern-.125emX}}
    
\begin{document}

\title[Localizing Software-Defined Metamaterials: Context Analysis]{Toward Localization in Terahertz-Operating Energy Harvesting Software-Defined Metamaterials: Context Analysis}

\subtitle{Invited Paper, Special Session on Teraherz Nanocommunication and Networking}

\author{Filip Lemic}
\affiliation{%
  \institution{Internet Technology and Data Science Lab (IDLab-Antwerp), University of Antwerpen - imec, Belgium}}
\email{filip.lemic@uantwerpen.be}

\author{Sergi Abadal}
\affiliation{%
  \institution{NaNoNetworking Center in Catalunya (N3Cat), Polytechnic University of Catalonia, Spain}}
\email{abadal@ac.upc.edu}

\author{Jeroen Famaey}
\affiliation{%
  \institution{Internet Technology and Data Science Lab (IDLab-Antwerp), University of Antwerpen - imec, Belgium}}
\email{jeroen.famaey@uantwerpen.be}

\renewcommand{\shortauthors}{F. Lemic, S. Abadal, J. Famaey}

\begin{abstract}

Software-defined metamaterials (SDMs) represent a novel paradigm for real-time control of metamaterials. 
SDMs are envisioned to enable a variety of exciting applications in the domains such as smart textiles and sensing in challenging conditions.    
Many of these applications envisage deformations of the SDM structure (e.g., rolling, bending, stretching).
This affects the relative position of the metamaterial elements and requires their localization relative to each other.  
The question of how to perform such localization is, however, yet to spark in the community.  
We consider that the metamaterial elements are controlled wirelessly through a Terahertz (THz)-operating nanonetwork.
Moreover, we consider the elements to be energy constrained, with their sole powering option being to harvest environmental energy.
For such a setup, we demonstrate sub-millimeter accuracy of the two-way Time of Flight (ToF)-based localization, as well as high availability of the service (i.e., consistently more than 80\% of the time), which is a result of the low energy consumed in localization.
Finally, we provide the localization context for a number of relevant system parameters such as operational frequency, bandwidth, and harvesting rate.
\vspace{-1mm}
\end{abstract}



\maketitle

\input{acronym_def}
\input{introduction}

\input{localization}

\input{results}

\input{conclusion}

\section*{Acknowledgments}

The author Filip Lemic was supported by the EU Marie Curie Actions Individual Fellowship project Scalable Localization-enabled In-body Terahertz Nanonetwork (SCaLeITN), grant nr. 893760. In addition, this work received support from the University of Antwerp's University Research Fund (BOF).
This work also received funding from the European Union via the Horizon 2020 Future Emerging Topics call (FETOPEN), grant no. 736876.
Finally, the authors would like to thank Prof. Chong Han for the fruitful discussions that helped us to significantly improve the quality of this paper.

\end{document}

%% file: acronym_def.tex

\acrodef{SDMs}{Software-Defined Metamaterials}
\acrodef{SDM}{Software-Defined Metamaterial}
\acrodef{THz}{Terahertz}
\acrodef{UWB}{Ultra Wide-Band}
\acrodef{FPGA}{Field Programmable Gate Array}
\acrodef{ToF}{Time of Flight}
\acrodef{AoA}{Angle of Arrival}
\acrodef{RSS}{Received Signal Strength}
\acrodef{AP}{Anchor Point}
\acrodef{3D}{3-Dimensional}
\acrodef{TS-OOK}{Time-Spread ON-OFF Keying}
\acrodef{RF}{Radio Frequency}

%% file: introduction.tex

\section{Introduction}

Metamaterials are manufactured structures with engineered properties that will enable controlled manipulation (e.g., transmission, reflection, absorption) of electromagnetic waves~\cite{lemic2019survey,abadal2020programmable}. 
To enable the real-time control of metamaterial elements, Liaskos et al.~\cite{liaskos2018new} proposed \acp{SDM}. 
For supporting the programming and control, the SDM paradigm envisions embedding a communication nanonetwork of controllers within the metamaterial, with the roles of controllers being i) actuation of the metamaterial elements, ii) collection of sensed readings from the metamaterial elements, and/or iii) two-way communication of the actuation success or sensed readings between the metamaterial elements and the outside world~\cite{lemic2020idling}.

Currently, there are several technological candidates for developing the SDM communication nanonetwork, including both wired~\cite{abadal2020programmable} (e.g., \acp{FPGA}) and wireless solutions (e.g., graphene-based \ac{THz} nanonetworks~\cite{lemic2019assessing}).  
These candidates come with strengths and weaknesses, with the primary concern being the communication wiring and form-factor vs. reliability trade-off between wired and wireless solutions~\cite{abadal2020programmable}.
The question of which technological candidate to utilize is currently not fully understood and is a subject of ongoing research. 

SDMs are envisioned to be embedded in smart textiles for enabling features such as wireless touch~\cite{tian2019wireless}.
SDMs are also expected to be used in vehicular communication for enhancing ranging capabilities~\cite{tak2017metamaterial} or noise cancellation~\cite{sato2007metamaterials}. 
Moreover, there are SDM-based applications that envision metamaterial-sensing on rough mobile surfaces (e.g., temperature~\cite{kairm2014concept} or strain sensing~\cite{melik2009flexible}) or in fluids~\cite{labidi2011meta}.
To enable such applications, one of the aims of metamaterial-focused research is to develop flexible SDMs with the possibility of bending, stretching, rolling, etc.~\cite{walia2015flexible}.    
Hence, the positions of the metamaterial elements relative to each other will be dynamically alterable.
Intuitively, for such SDM-based applications the location of a given metamaterial element will play a role in the control of that element.   
However and to the best of our knowledge, the question on how to localize the metamaterial elements relative to each other did not receive any attention in the community to date.

As one of the first steps toward addressing this issue, we consider the usage of a THz nanonetwork in \acp{SDM}. 
We do that for a case of highly energy-constrained SDMs, in which harvesting the environmental energy is the sole powering option for the metamaterial elements.
This is done as it is expected that such energy harvesting SDMs will mostly be utilized in the future, primarily to reduce the form-factor of SDMs by removing the need for wiring~\cite{lemic2019survey}. 
Borrowing the knowledge of traditional localization utilizing \ac{RF} signals, we demonstrate the feasibility of localizing SDM elements in the above-described setup by utilizing THz frequencies (i.e., 300 GHz to 10 THz). 
Specifically, we show a sub-millimeter average localization error of a THz-based two-way \ac{ToF} localization approach.
In addition, we demonstrate consistently high availability of the localization service, resulting from the low energy consumed in the process.  
Finally, we characterize the influence of several relevant system parameters (e.g., operational frequency, bandwidth, energy harvesting rate, SDM spacing) on the accuracy and availability of localization.

%% file: localization.tex
\vspace{-1.6mm}
\section{Context Analysis}
\subsection{Software-defined Metamaterials} 

\begin{figure}[t]
\vspace{-1mm}
\centering
\includegraphics[width=\linewidth]{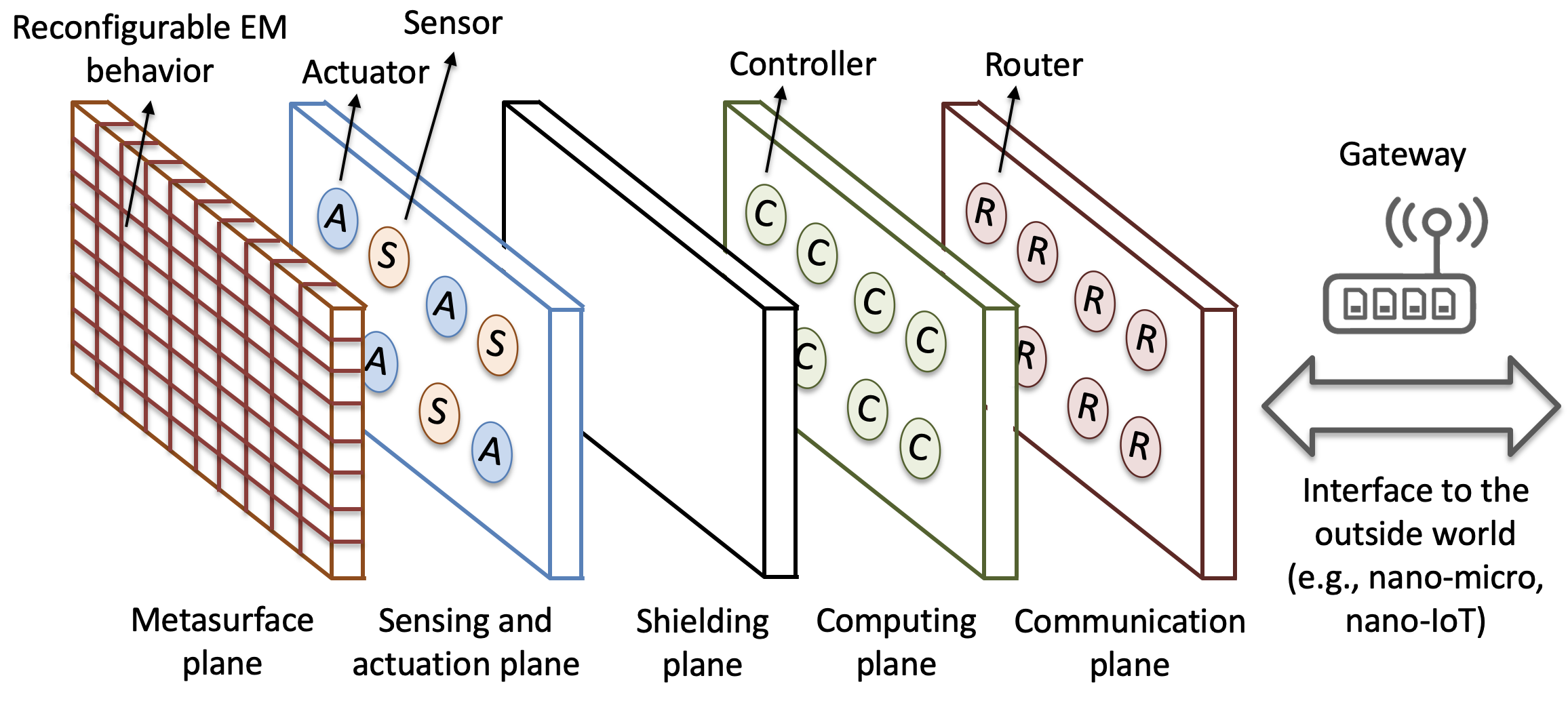}
\vspace{-6mm}
\caption{SDM paradigm for controlling metamaterials~\cite{lemic2020idling}}
\label{fig:architecture}
\vspace{-2mm}
\end{figure}

Figure~\ref{fig:architecture} depicts the usual SDM architecture, where each controller can interact with its associated unit cells. 
Each unit cell represents a sensor and/or actuator communicating with and being controlled by a given set of controllers. 
These sensors and actuators are used as an abstraction for functionalities of adjusting the properties and delivering the readings of the metamaterial elements. 
Moreover, in Figure 1, the routing plane is used for distributing the desired behavior of the metamaterials across the controllers and for enabling the communication between the controllers and the outside world. 
We consider THz-based wireless nanocommunication between the controllers and the sensors and actuators, which allows for a tiny form factor and small energy consumption~\cite{jornet2013graphene}. 
Moreover, we use \ac{TS-OOK} as a modulation and coding scheme as it is a \textit{de-facto} standard for THz nanocommunication~\cite{jornet2011information}. 

\subsection{Energy Harvesting}

\begin{figure}[t]
\vspace{-1mm}
\includegraphics[width=0.95\linewidth]{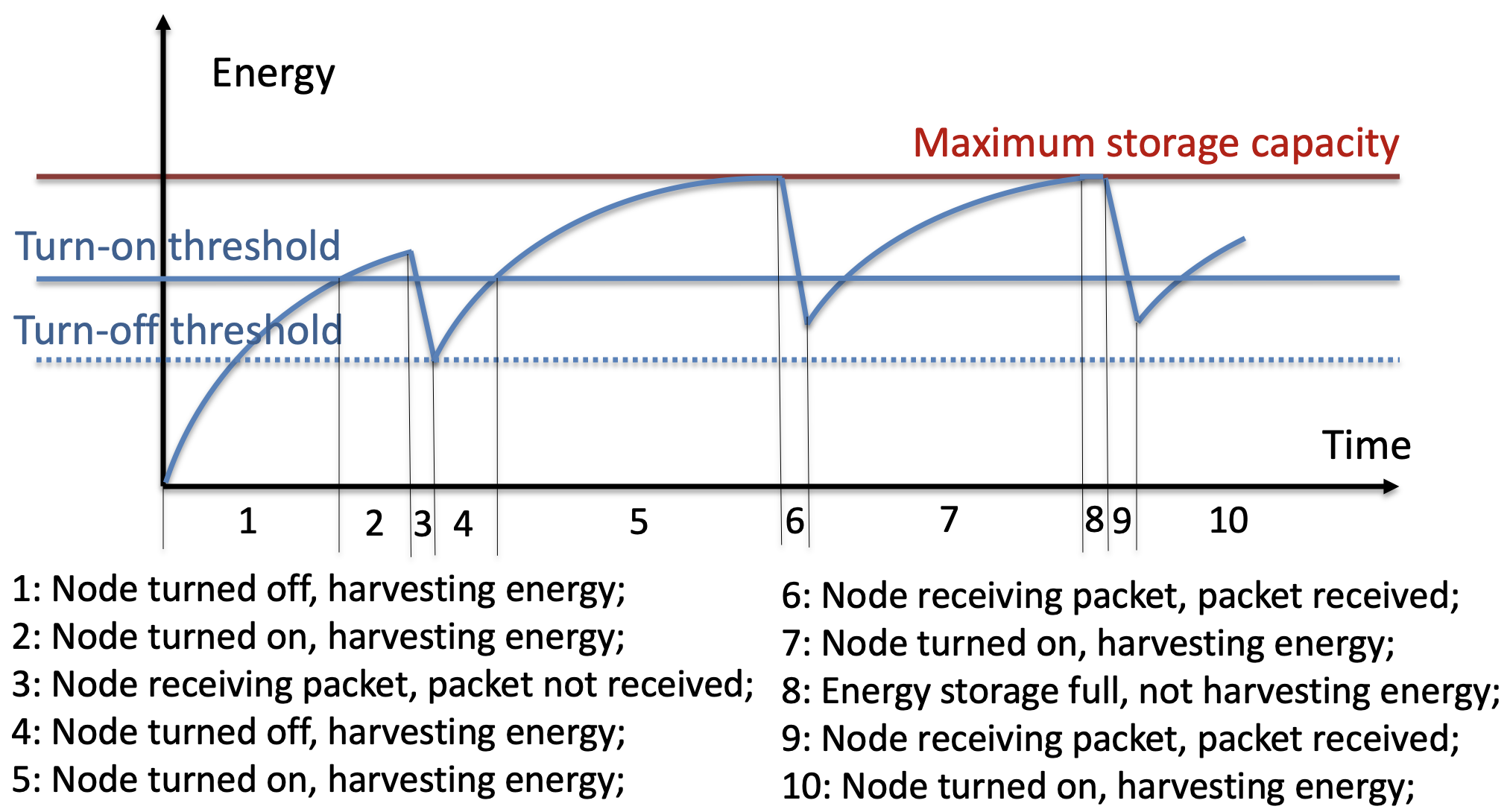}
\vspace{-3mm}
\caption{Lifecycle of an energy harvesting nanonode~\cite{lemic2019assessing}}
\label{fig:intermittent}
\vspace{-7mm}
\end{figure}

The sensors and actuators can be considered as battery-less energy harvesting nanonodes. 
The usual energy lifecycle of such a nanonode is depicted in Figure~\ref{fig:intermittent}.
At certain points in time (i.e., the ``turn-off threshold'') the energy of the nanonode will be critically low, thus the nanonode will not be able to operate. 
At a certain later point, the nanonode will have harvested enough energy to turn on again (i.e., ``turn-on threshold'') and will become operational again.  
Intuitively, the nanonode will continue to harvest energy if it is turned on until its energy level reaches the maximum storage capacity.
During transmission, reception, or any other operational periods (e.g., idling, sensing, actuation), the nanonode will lose certain amounts of energy, while at the same time gaining some amount of energy due to harvesting.

The current state-of-the-art nanoscale energy harvesters the exploit piezoelectric effect of ZnO nanowires~\cite{wang2008towards}, where the energy is harvested in nanowires' compress-and-release cycles.
The harvested energy can be specified with the duration of the harvesting cycle $t_{cycle}$ and the harvested charge per cycle $\Delta Q$.
Energy harvesting can be accurately modeled as an exponential process~\cite{jornet2012joint}, accounting for the total capacitance $C_{cap}$ of the nanonode, where $C_{cap} = 2 E_{max} / V_g^2$, i.e., $C_{cap}$ depends on the maximum energy storage capacity $E_{max}$ and the generator voltage $V_g$.  
In the modeling, it is required to know in which harvesting cycle $n_{cycle}$ the nanonode is, given its current energy level $E_{n_{cycle}}$, which can be derived from~\cite{jornet2012joint} as follows:

\begin{equation}
\label{eq1}
n_{cycle} = \ceil*{\frac{- V_g C_{cap}}{\Delta Q} ln\left(1 - \sqrt{\frac{2 E_{n_{cycle}}}{C_{cap} V_g^2}}\right)}.
\end{equation}

The energy in the next energy cycle $n_{cycle} + 1$ is then: 

\begin{equation}
\label{eq2}
E_{n_{cycle+1}} = \frac{C_{cap} V_g^2}{2} \left(1- e^{-\frac{\Delta Q (n_{cycle} + 1)}{V_g C_{cap}}}\right)^2.
\end{equation}
\vspace{-3mm}

\subsection{Localizing Software-Defined Metamaterials}

Traditional \ac{RF} localization approaches can be categorized into fingerprinting, proximity, and geometric-based.
Fingerprinting approaches rely on correlating various signal features from an unknown location with a set of respective signal features at presurveyed locations~\cite{caso2019vifi}. 
Due to the fact that it is practically impossible to presurvey localization at a nanoscale, we believe fingerprinting is infeasible for localization in SDMs. 
Proximity-based approaches rely on the proximity of a device to be localized and a set of anchors with known locations. 
These approaches are designed to provide coarse-grained localization~\cite{lemic2015experimental} and are, therefore, not suitable for the desired nanoscale localization with very high accuracy.
Geometrical approaches utilize different signal features (mostly \ac{RSS}, \ac{AoA}, and ToF) for estimating distances (or angles) between devices, which then serve as a basis for estimating the unknown locations~\cite{lemic2016localization}.

AoAs cannot be estimated without antenna arrays or complex signal processing, both being infeasible for the considered nanonodes with constrained capabilities~\cite{rong2006angle}.
RSS is a highly fluctuating signal feature with logarithmic dependence to the distance between devices, hence it is known to be highly inaccurate~\cite{lemic2016localization}. 
Estimation of the ToF requires tight synchronization between the devices, which is hardly achievable for the considered resource-constrained nanonodes.  
Luckily, a two-way ToF-based approach removes the synchronization requirement and we, therefore, reason that two-way ToF-based localization can potentially enable localization in SDMs.
For estimating two-way ToF, a signal is transmitted by one device and the time measuring is started. Upon reception, the other device transmits the signal back containing the estimate of the time passed between the reception of the original signal and transmission of the new one. Upon the following reception, the first device can estimate the two-way ToF by subtracting the time that passed between the reception of the original signal and the transmission of the new one from the total time since the original transmission.   

The accuracy of this method will depend on the sampling constraints resulting from the Nyquist theorem. 
Borrowing the classification from~\cite{xiong2015tonetrack}, in Table~\ref{tab:table1} we summarize the raw resolution of different physical layers prominently used for localization, with the raw resolution defined as the speed of light divided by the available bandwidth.
In the THz frequencies, at least 10~GHz spectrum windows are unlicensed and ubiquitously available~\cite{boronin2014capacity}, while at the nanoscale the available bandwidth increases to THz widths~\cite{abadal2019graphene}.
This respectively yields the achievable raw resolution of less than 3~cm and 0.3~mm at the macro- and nanoscale, which demonstrates a potential for high accuracy in THz-based localization compared to other physical layers used for localization.  

\begin{table}[t]
\vspace{-1mm}
\centering
\caption{Popular physical layers used in localization~\cite{xiong2015tonetrack}}
\label{tab:table1}
\small
\vspace{-2mm}
\begin{tabular}{p{2.9cm} p{2.0cm} p{2.2cm}}
\textbf{Physical layer}  & \hfil \textbf{Bandwidth}      & \hfil \textbf{Raw resolution} \\ \hline
IEEE~802.11a/g           & \hfil 20~MHz                  & \hfil 15~m                    \\
IEEE~802.11n             & \hfil 40~MHz                  & \hfil 7.5~m                   \\
IEEE~802.11ac            & \hfil $<$160~MHz              & \hfil $>$1.9~m                \\
Ultra WideBand (UWB)     & \hfil $>$500~MHz              & \hfil $<$0.6~m                \\ 
IEEE~802.11ad            & \hfil $>$2~GHz                & \hfil $<$15~cm   			 \\ 
\textbf{Terahertz - macroscale} & \hfil $\bm{>}$\textbf{10~GHz} & \hfil $\bm{<}$\textbf{3~cm}   \\ 
\textbf{Terahertz - nanoscale} & \hfil $\bm{>}$\textbf{1~THz} & \hfil $\bm{<}$\textbf{0.3~mm}   \\ \hline
\end{tabular}
\vspace{-3mm}
\end{table}

\begin{figure}[t]
\vspace{-1mm}
\centering
\includegraphics[width=0.64\linewidth]{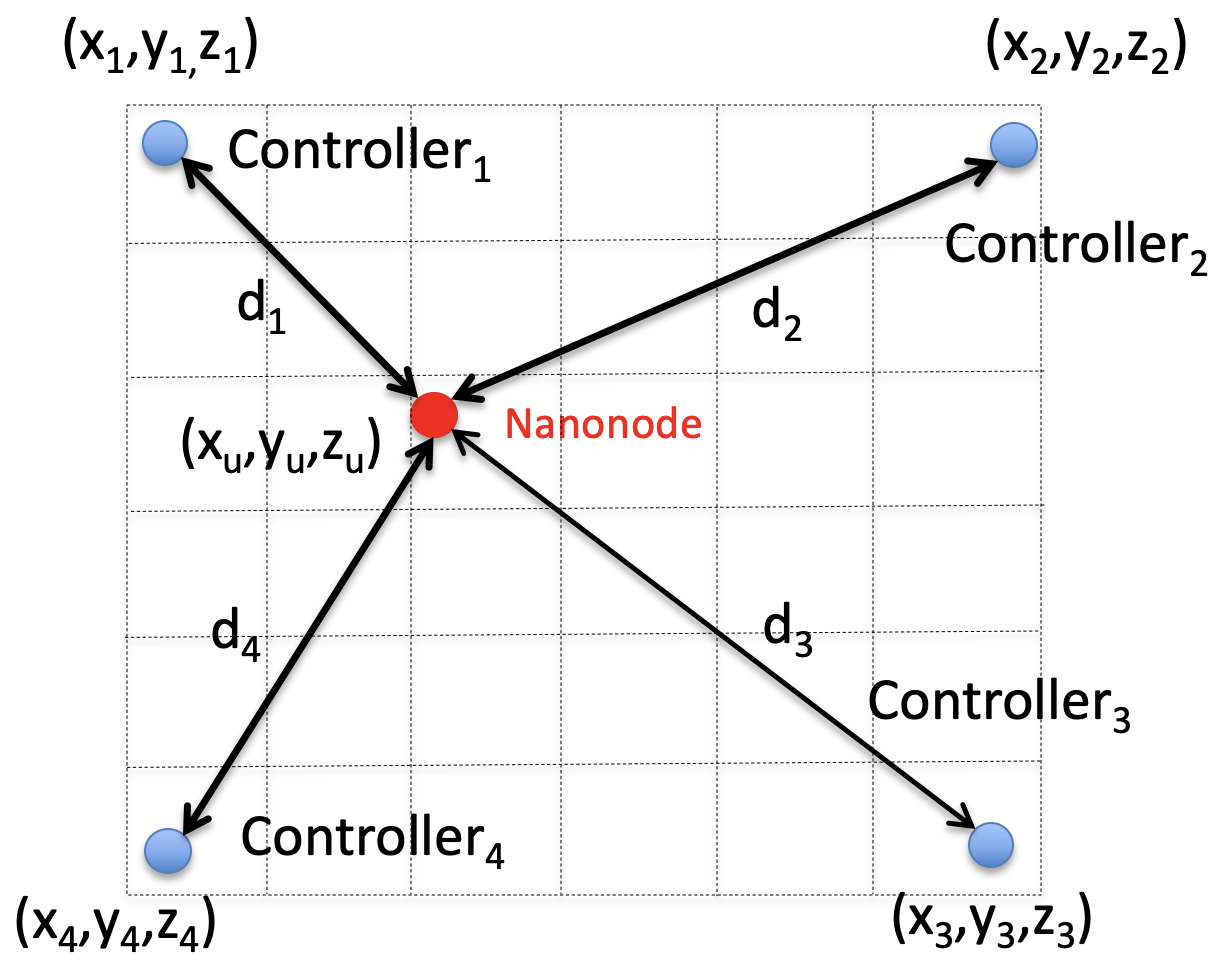}
\vspace{-4mm}
\caption{Trilateration setup}
\label{fig:localization_setup}
\vspace{-5mm}
\end{figure}  

We envision the localization setup as depicted in Figure~\ref{fig:localization_setup}.
Specifically, there are four controllers that are not energy-constrained and whose locations are known. 
The assumption of the controllers not being energy constrained (i.e., mains- or battery-powered) has been widely used in the literature (e.g.,~\cite{lemic2019assessing,lemic2020idling}).
The assumption that their locations are known is reasonable as their locations can be fixed and measured (e.g., when an SDM is mounted on a rough surface).
Adversely, their locations can be estimated by utilizing some of the traditional approaches, as they are not size-, resource-, or energy-constrained.  
Localization could also be done in two-steps, where first the controllers are localized by utilizing two-way ToF-based trilateration, followed by localizing the nanonodes. 
This can be done as the controllers are not energy-constrained, thus they can utilize higher transmit power for localization, which in turn would attribute to the enhanced range of localization.

Under the above assumptions, the localization process includes each controller transmitting a TS-OOK pulse, which is upon reception retransmitted by the nanonode whose location is to be estimated.
If the retransmitted signals are received by all four controllers, the distances between the controllers and the nanonode can be estimated from the two-way ToF measurements.
The reasons why the retransmitted signals could be missed are two-fold. 
First, the nanonode's energy could be depleted, hence it would not be able to retransmit the original signal(s).
Second, due to mobility (e.g., stretching of an SDM) a controller could be out of the communicating range.   
If the distances between the controllers and the nanonode can be estimated, a standard trilateration approach can the be utilized for estimating the unknown location in a \ac{3D} space.  
More details on how to estimate the distance between each controller and the nanonode, as well as the basics of trilateration, can be found in e.g.,~\cite{karl2007protocols}.

\begin{figure}[t]
\vspace{-1mm}
\centering
\includegraphics[width=0.78\linewidth]{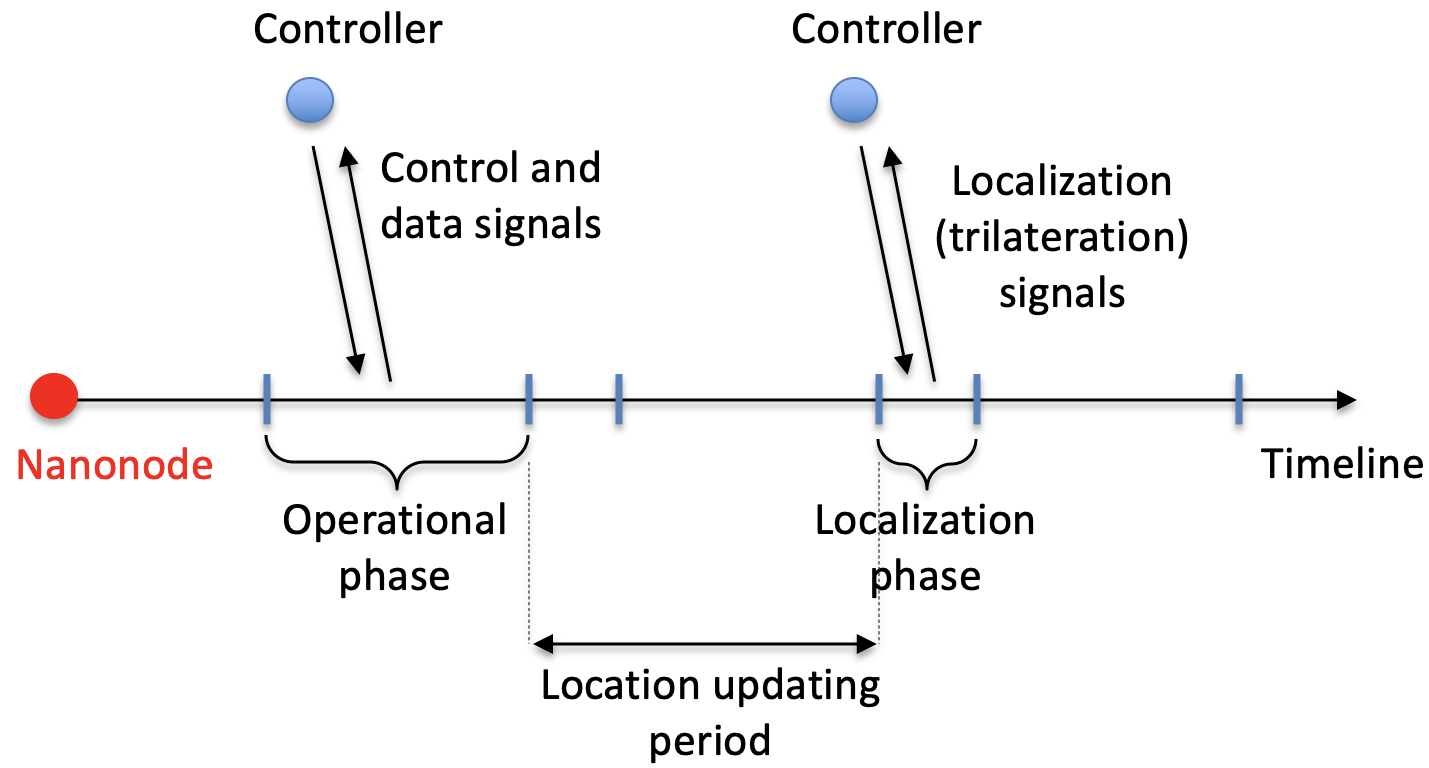}
\vspace{-4mm}
\caption{Operational timeline of a nanonode}
\label{fig:operational_timeline}
\vspace{-5mm}
\end{figure}

The operational timeline of the nanonode whose location is to be estimated is given in Figure~\ref{fig:operational_timeline}. 
Due to potentially continuous mobility, the location of the nanonode has to be estimated periodically. 
Hence, we envision two phases in the operational timeline, i.e., the operational phase during which the nanonode performs its function (e.g., sensing, actuation) and the localization phase during which its location is being estimated.   
The frequency of location estimation in this scenario is application-specific and depends on the location updating period, as shown in the figure. 

%% file: results.tex

\section{Evaluation}
\label{results}

\subsection{Evaluation Methodology}
 
The goal of the evaluation is to establish the accuracy and availability of the THz-based two-way \ac{ToF} approach for localization in SDMs.  
These metrics are to be derived as functions of several relevant system parameters, i.e., operational frequency, bandwidth, energy harvesting rate, location update period, spacing between metamaterial elements, and receiver sensitivity.  
The accuracy is characterized by the localization error, which is defined as the Euclidean distance between the true and estimated location of a nanonode.  
The availability is defined as the ratio between the number of successful and the total number of localization attempts.
The summary of the default simulation parameters is given in Table~\ref{tab:paramaters}. 

\begin{table}[t]
\vspace{-2mm}
\small
\begin{center}
\caption{Simulation parameters}
\vspace{-2mm}
\label{tab:paramaters}
\begin{tabular}{l r}
\hline
\textbf{Parameter} & \textbf{Value} \\
\hline
Number of nanonodes & (25x25) 625 \\
Distance between nanonodes [mm] & 0.9 \\
Generator voltage $V_g$ [V] & 0.42 \\
Transmit power $P_{T_X}$ [dBm] & -20 \\
Energy consumed in pulse reception $E_{R_{X pulse}}$ [pJ] & 0.1 \\
Energy consumed in pulse transmission $E_{T_{X pulse}}$ [pJ] & 1.0 \\
Data packet size [bits] & 8 \\
Maximum energy storage capacity [pJ] & 800 \\
Turn OFF/ON thresholds [pJ] & 10/0 \\
Simulation time [\# of iterations] & 1000 \\
Harvesting cycle duration [ms] & 20 \\
Harvested charge per cycle [pC] & 6 \\
Localization update period [sec] & 0.1 \\
Operational bandwidth [THz] & 1 \\ 
Receiver sensitivity [dBm] & -100 \\ 
Operational frequency [THz] & 1 \\ 
\hline
\end{tabular}
\end{center}
\vspace{-5mm}
\end{table}

In an in-house made Python-based simulator, we define a set of 625 (25$x$25) nanonodes in a grid-like fashion.
The default distance between neighboring nanonodes is set to 0.9~mm, which correlates to spacing needed for controlling electromagnetic waves in the mmWave frequencies (i.e., with the operational frequency of 250~GHz, assuming the spacing equals to 3/4 of the wavelength), which is among the most exciting SDM applications~\cite{abadal2020programmable}.
The four nanonodes in the corners of the grid are given the role of controllers.
Hence, they are not energy constrained and their locations are known and fixed.      
The other nanonodes are powered through energy harvesting, with their energy levels being modeled by Equations (1) and (2). 
Moreover, we assume that the energy is harvested from air-vibrations, hence we specify the default harvesting cycle duration and harvested charge per cycle of 20~ms and 6~pC~\cite{jornet2012joint}, respectively.  
The energy consumed in transmitting ($E_{T_{X pulse}}$) and receiving ($E_{R_{X pulse}}$) a TS-OOK pulse are set to 1 and 0.1~pJ~\cite{jornet2012joint,hossain2018terasim}, receptively.
Furthermore, the transmit power is set to -20~dBm, which is again in line with the existing literature~\cite{jornet2012joint,hossain2018terasim}. 
The energy is assumed to be consumed both during the localization and operational phases of a nanonode. 
During the localization phase, the energy can be consumed in the reception and retransmission of a TS-OOK pulse for deriving the two-way ToF.
In the operational phase, the energy is consumed in the reception of an 8 bits long packet, with the bits of the packet (i.e., logical ``0''s and ``1''s) being drawn from a uniform distribution.    
The reception of such a packet models the control of a nanonode for which an 8 bits long packet is usually used~\cite{lemic2019survey,abadal2020programmable}.  
The default operational bandwidth is set to 10~GHz, as there are several windows that can provide such (or more) bandwidth at THz frequencies~\cite{boronin2014capacity}. 

The locations of the nanonodes are selected randomly in the area of sizes $(x,y,z)=(d,d,d/2)$, with d being the distance between two controllers on the same edge.
This random selection of locations inside a bounded area resembles a scenario in which a flexible SDM is attached to an uneven convex surface, although we acknowledge additional constraints in the localization area are possible.   
Two-way ToF measurements are derived from the true distances between a nanonode and the controllers, to which a zero-mean Gaussian-drawn variability is added.   
The standard deviation of the variability is derived as the ratio between the speed of light and bandwidth (i.e., the raw resolution), which is an often utilized method for simulating the ToF variability~\cite{wirstrom2015localization,lemic2016localization}.

Location estimates can be established if the two-way ToF measurements between a given nanonode and all 4 controllers are obtained.
This is characterized by comparing the strength of the received signal with the receiver sensitivity, i.e., if the received signal strength is higher than the receiver sensitivity, the signal is considered as received.
The received signal strength $P_{Rx}$ is obtained by subtracting absorption and spreading losses from the transmit power $P_{Tx}$, which is a THz nanoscale channel modeling method often used in the literature (e.g.,~\cite{jornet2011channel,kokkoniemi2014frequency}).
The received signal strength is given as follows, with $d$, $f$, and $c$ being respectively the distance between devices, operating frequency, and speed of light. Moreover, $k(f)$ is a frequency-dependent medium absorption coefficient, with its values obtained from the HITRAN database~\cite{rothman1987hitran}:

\vspace{-3mm}
\begin{equation}
P_{Rx} [dBm] = P_{Tx} - k(f) d 10 log_{10}(e) - 20 log\left(\frac{4 \pi f d}{c}\right).
\end{equation}
\vspace{-3mm}
\subsection{Evaluation Results}

\begin{figure}[t]
\vspace{-1mm}
\centering
\includegraphics[width=0.86\linewidth]{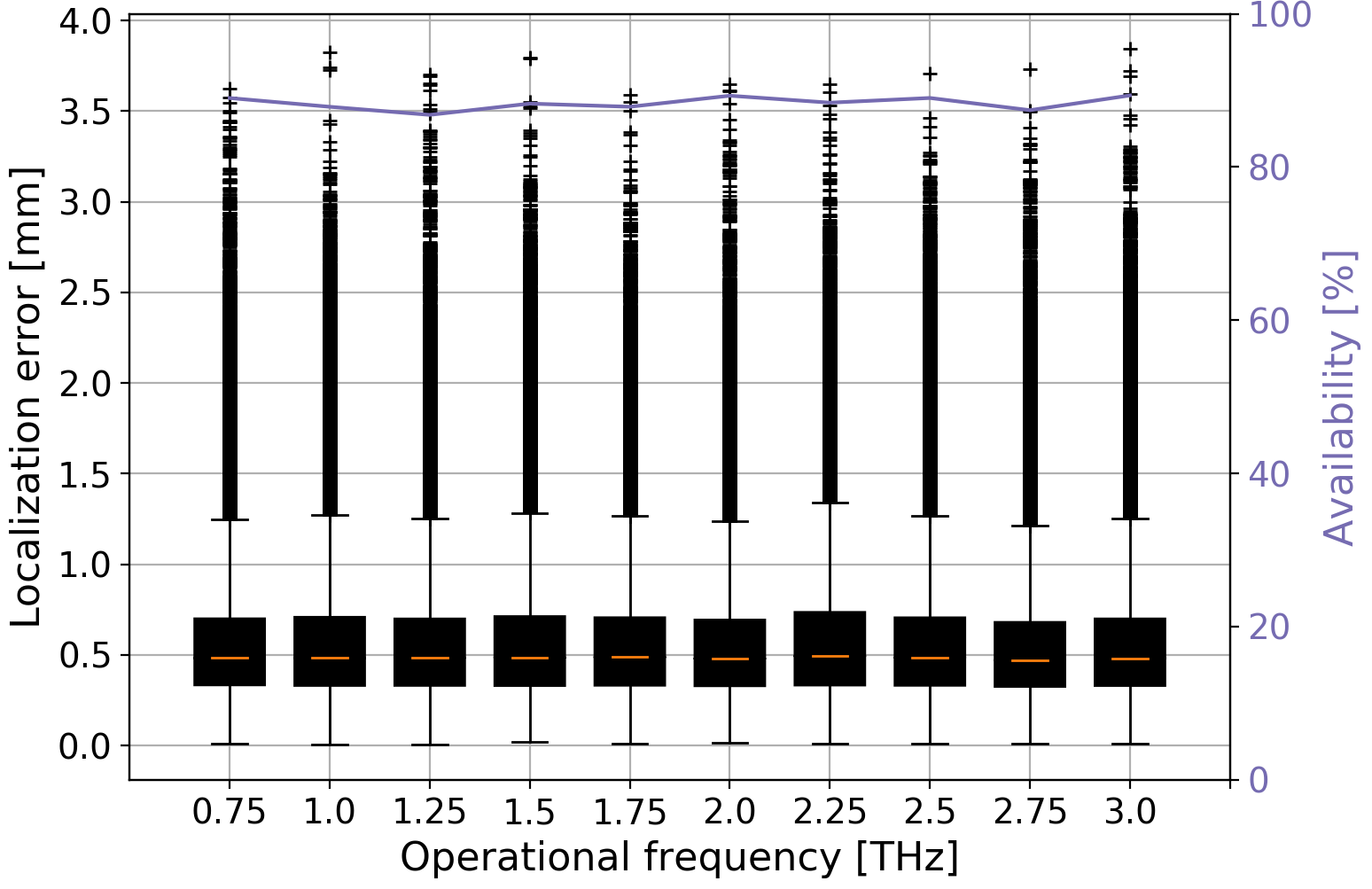}
\vspace{-3mm}
\caption{Accuracy and availability vs. operational frequency}
\label{fig:op_frequency}
\vspace{-1mm}
\end{figure}

Figure~\ref{fig:op_frequency} depicts the accuracy and availability of the localization service as a function of operating frequency. 
As visible, the operating frequency does not significantly affect the specified metrics.
This is because all the nanonodes are in the default setup in the range of all controllers, regardless of the operating frequency. 
Thus, the two-way ToF measurements between all nanonodes and all controllers and can be obtained and localization can be performed.   
In terms of accuracy, the localization service achieves around 0.5~mm in average localization error, which we believe demonstrates the feasibility of the two-way ToF-based trilateration in SDMs. 
Moreover, the outliers (90$^{th}$ percentile of all estimates) experience localization errors of less than 4.0~mm, which again shows the promise of the two-way ToF-based trilateration service.
Finally, the availability of the service is continuously around 90\%, with the only reason for not being able to estimate the location resulting from the depletion of the available energy.  

\begin{figure}[t]
\vspace{-1mm}
\centering
\includegraphics[width=0.86\linewidth]{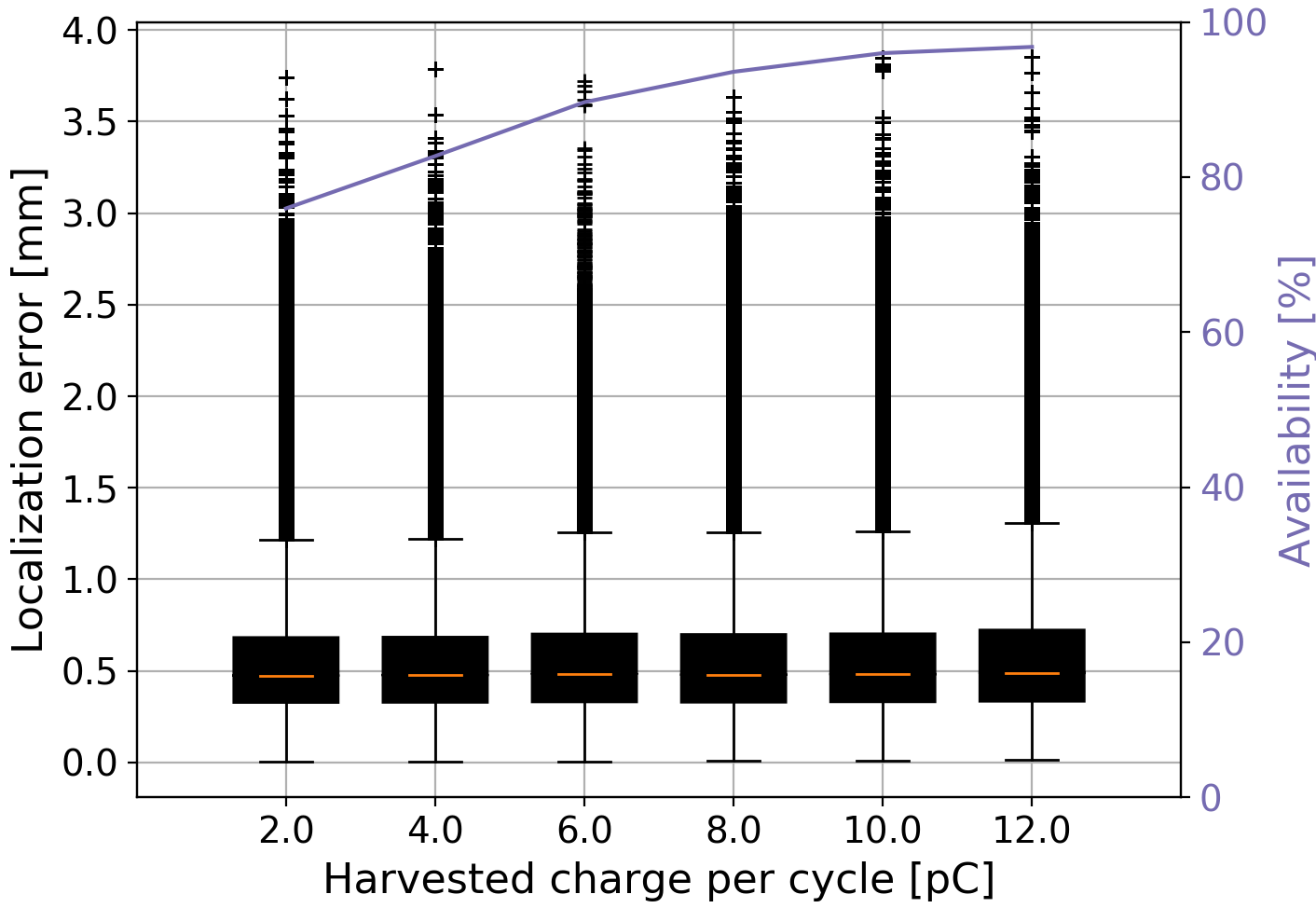}
\vspace{-3mm}
\caption{Accuracy and availability vs. energy harvesting rate}
\label{fig:harvesting}
\vspace{-3mm}
\end{figure}

\begin{figure}[t]
\vspace{-1mm}
\centering
\includegraphics[width=0.86\linewidth]{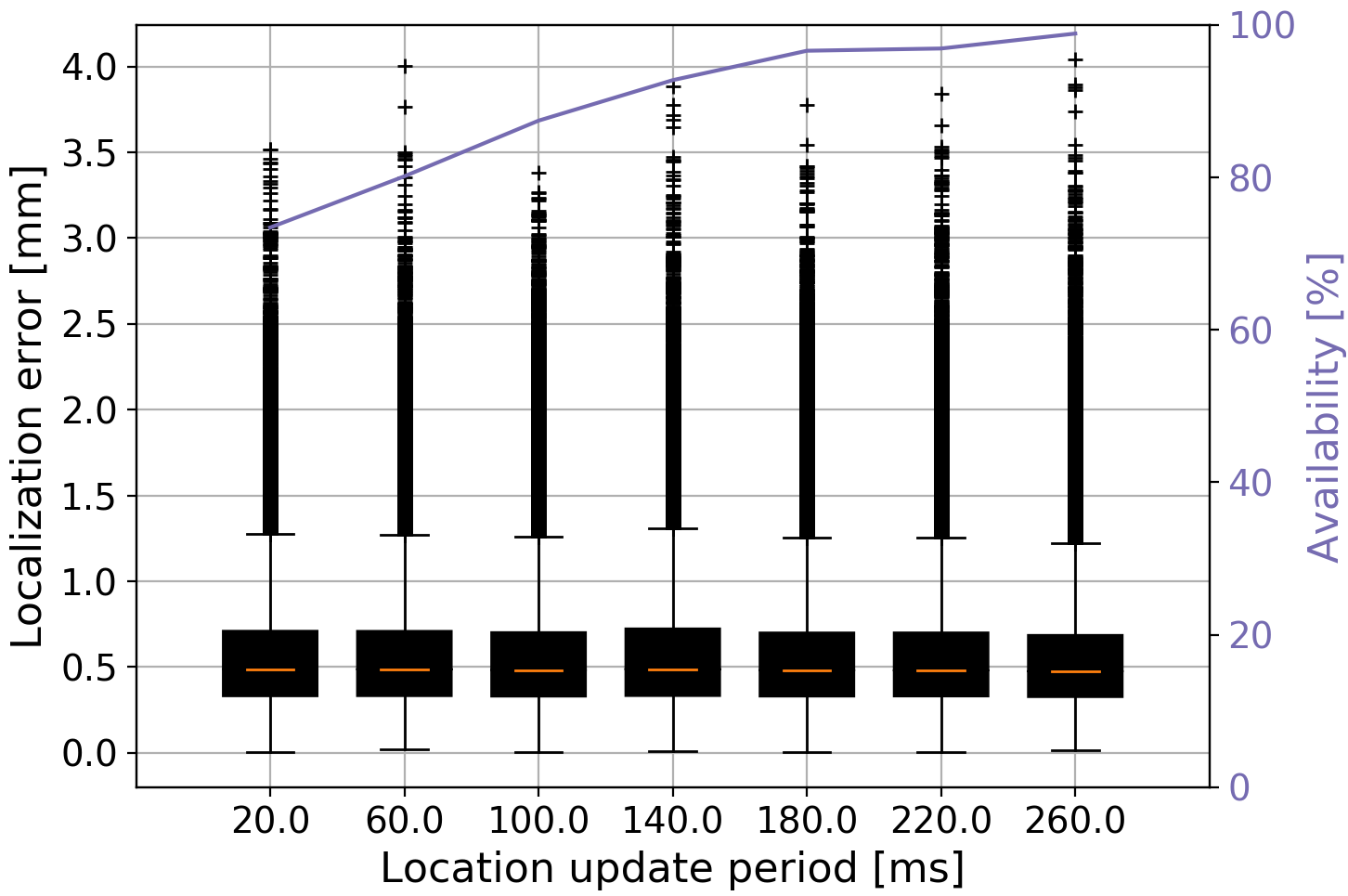}
\vspace{-3mm}
\caption{Accuracy and availability vs. location update period}
\label{fig:loc_frequency}
\vspace{-1mm}
\end{figure}

Similar conclusions can be made for the accuracy of localization as a function of energy harvesting rate (Figure~\ref{fig:harvesting}) and location update period (Figure~\ref{fig:loc_frequency}).  
This is because these parameters only affect the probability of generating two-way ToF measurements, and not their quality.
Thus, if these measurements are obtained, location can be estimated with a consistently high accuracy.   
However, as the energy harvesting rate or location update period increase, the availability of the localization service is enhanced.
For example, if the harvested charge is increased from 2 to 10~pJ per cycle, the availability is increased by roughly 10\%.
Similarly, if the localization update period increases from 20 to 220~ms, the availability increases from 80 to 100\%.
The reason for that lies in the fact that higher harvesting rate and location update period represent more harvested energy and less consumed energy, respectively.
Therefore, the nanonodes more often have sufficient energy to generate two-way ToF measurements and consequently estimate the location, which enhances the availability of the localization service.   

\begin{figure}[t]
\vspace{-1mm}
\centering
\includegraphics[width=0.86\linewidth]{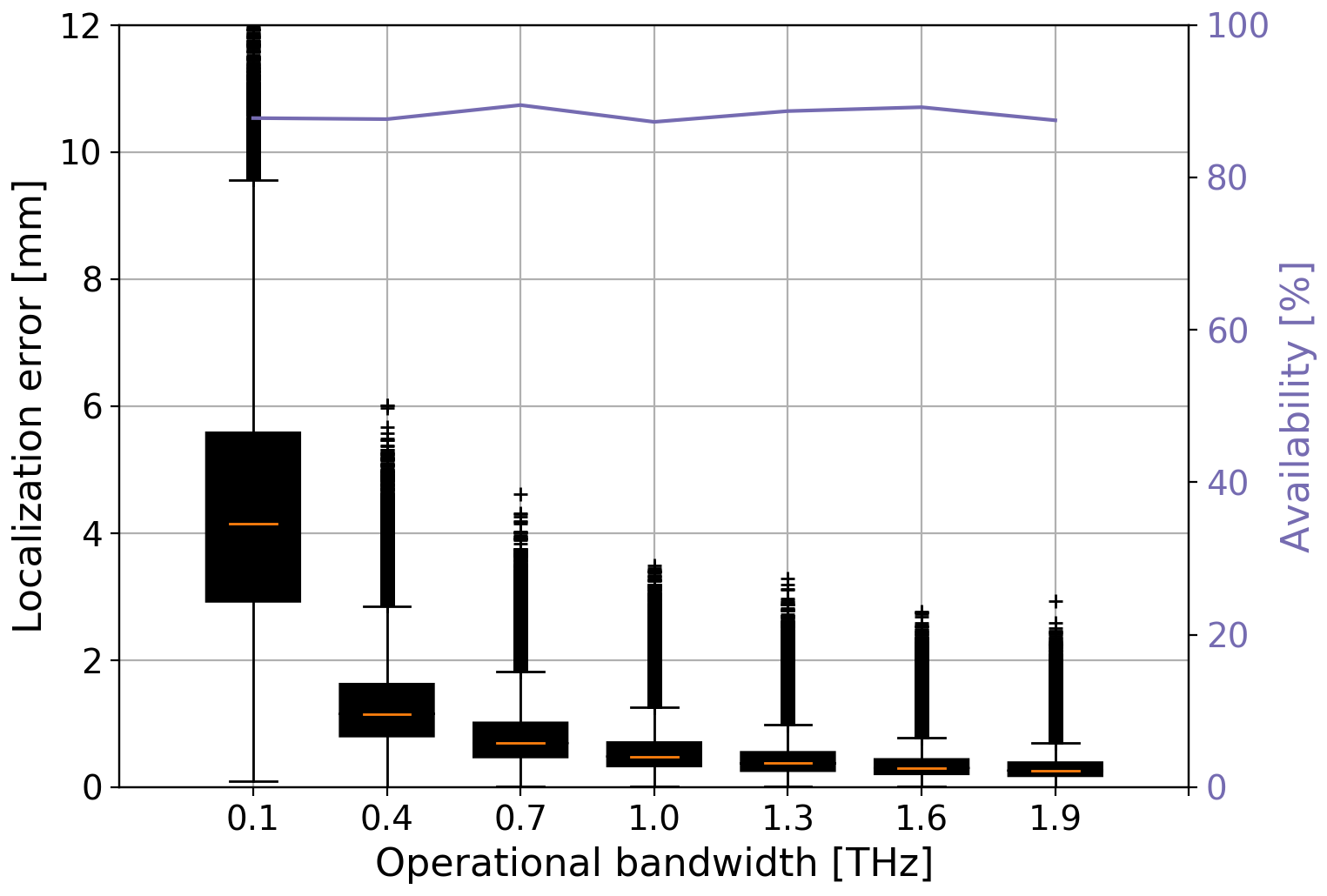}
\vspace{-3mm}
\caption{Accuracy and availability vs. operational bandwidth}
\label{fig:bandwidth}
\vspace{-1mm}
\end{figure}

\begin{figure}[t]
\vspace{-1mm}
\centering
\includegraphics[width=0.86\linewidth]{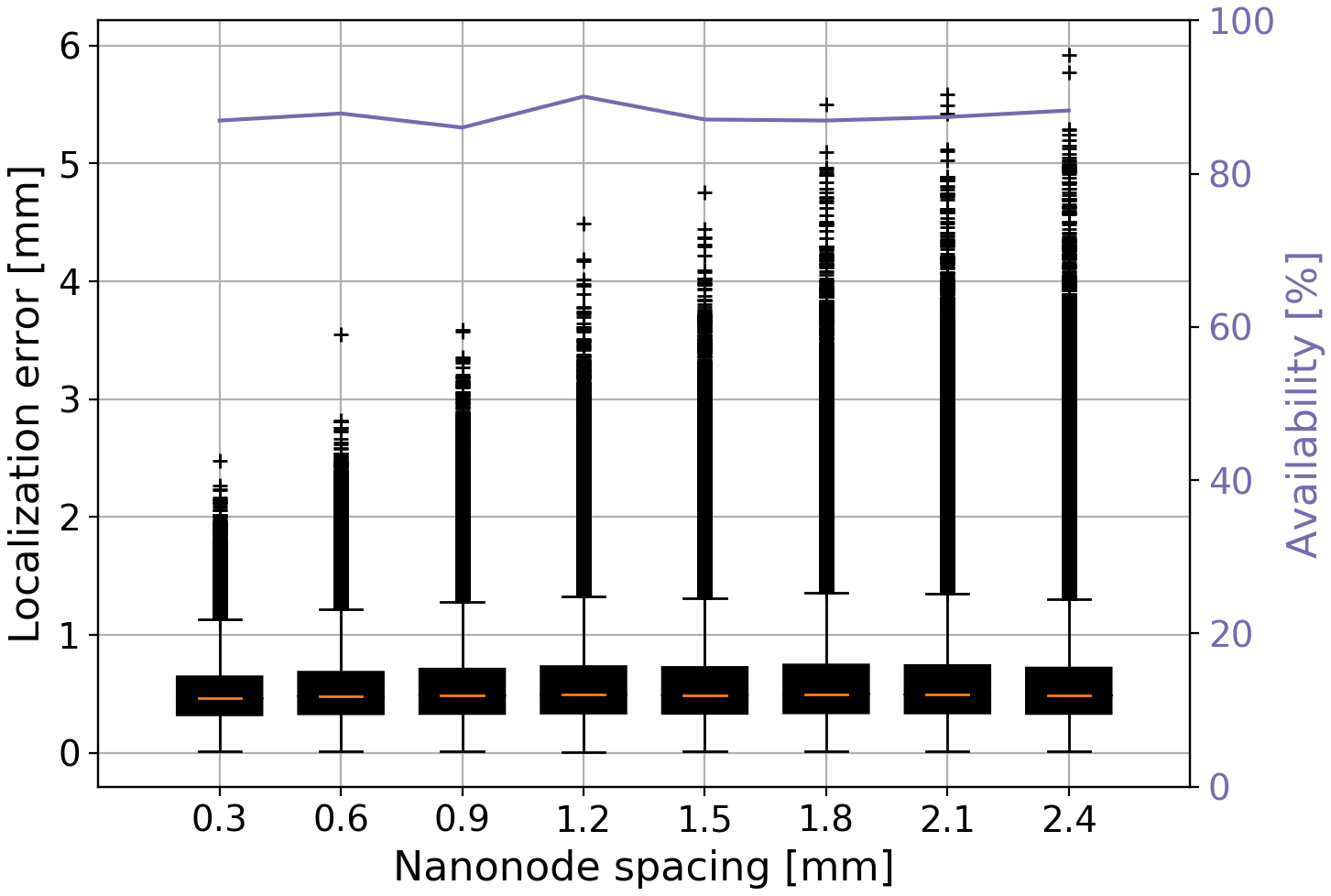}
\vspace{-3mm}
\caption{Accuracy and availability vs. spacing between nanonodes}
\label{fig:spacing}
\vspace{-3mm}
\end{figure}

Figures~\ref{fig:bandwidth} and~\ref{fig:spacing} depict the localization accuracy and availability in relation to the utilized bandwidth and spacing between nanonodes.  
As visible from the figures, both parameters have an impact on the localization accuracy, while not influencing the availability of the service.
This is because the increase in the utilized bandwidth increases the signal sampling rate.
This in turn reduces the variability of the ToF measurements, benefiting the accuracy of localization.
For example, if the utilized bandwidth is increased from 100~GHz to 1~THz, the average localization error is reduced from roughly 4.0 to less than 0.5~mm.   
Similarly, the spacing between the nanonodes affects the localization accuracy, however in this case only the outliers are affected.
This is because as the spacing increases, the sizes of the localization area increase as well, which in turn results in larger errors of the outliers.
This is a well-known behavior in more traditional localization approaches, e.g.,~\cite{behboodi2017hypothesis,lemic2016toward}. 

\begin{figure}[t]
\vspace{-2mm}
\centering
\subfigure[0.9~mm spacing between neighboring nanonodes\vspace{-4mm}]{\includegraphics[width=0.86\linewidth]{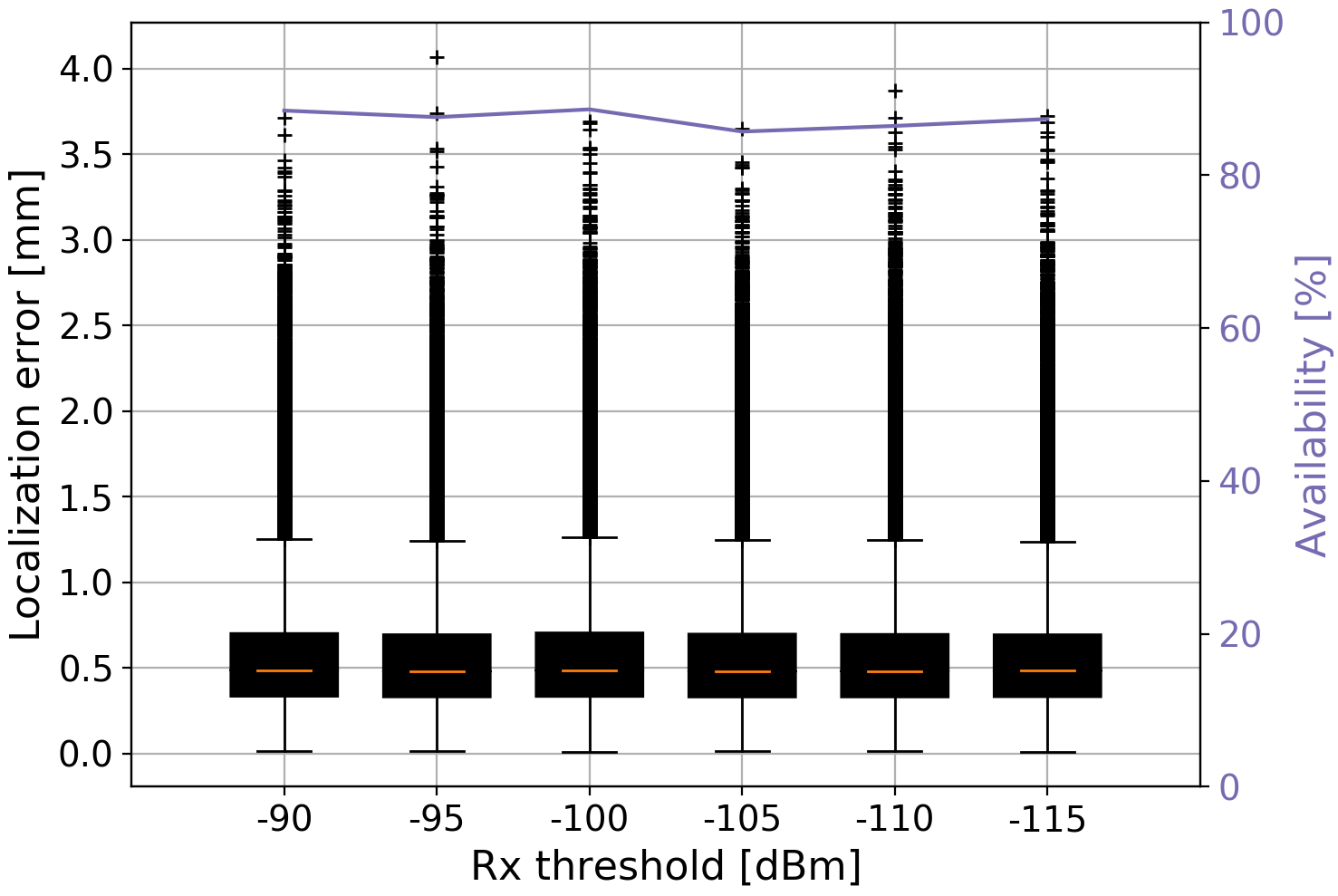}}
\subfigure[3~mm spacing between neighboring nanonodes]{\vspace{-4mm}\includegraphics[width=0.84\linewidth]{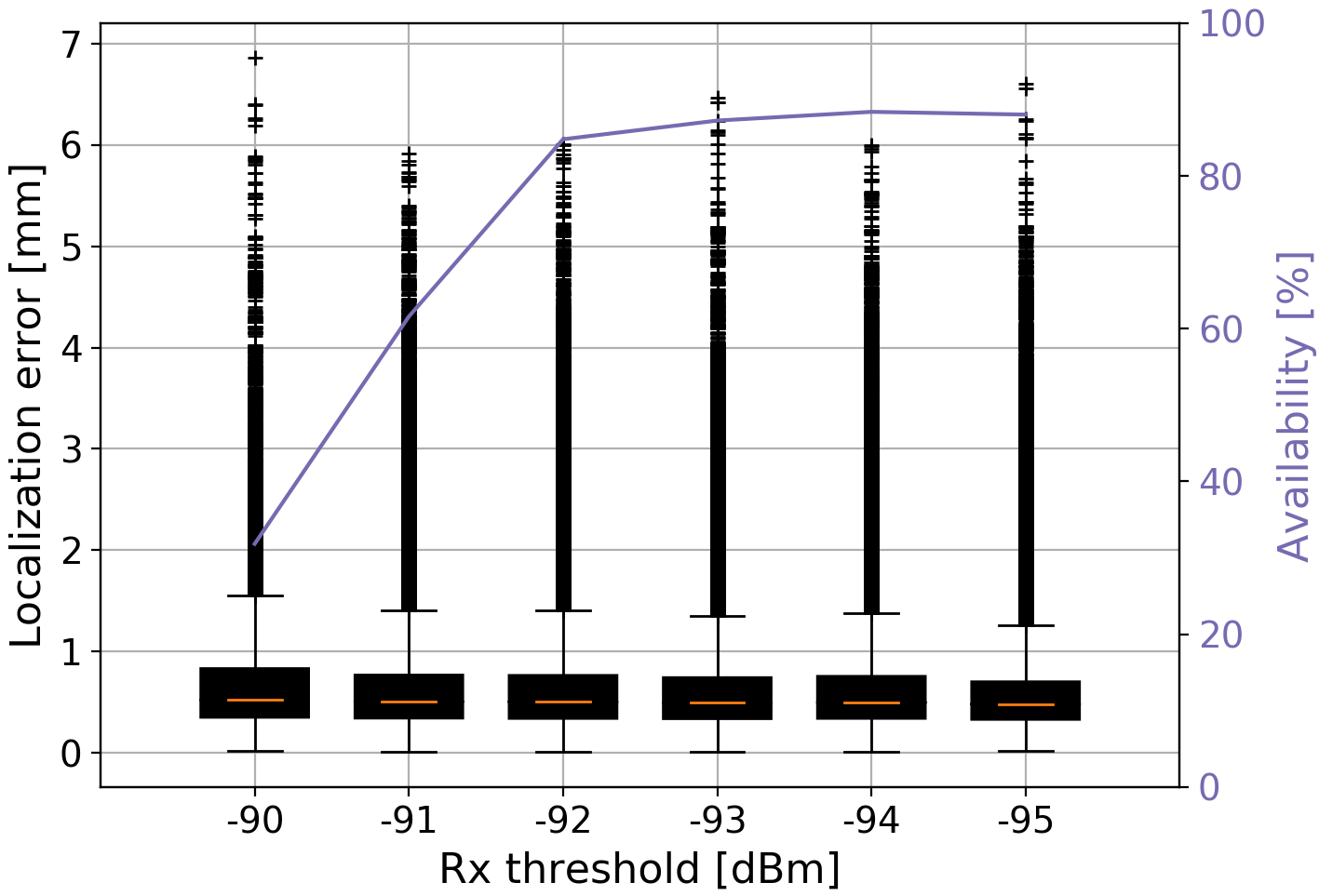}}
\vspace{-4mm}
\caption{Accuracy and availability vs. receiver sensitivity}
\label{fig:rx_sensitivity}
\vspace{-5mm}
\end{figure}

Finally, as depicted in Figure~\ref{fig:rx_sensitivity}a), the receiver sensitivity does not affect the localization accuracy and availability in the default setup.
This is because all the nanonodes are in the range of the controllers and experience relatively high received power.
The effect on the accuracy and availability is only visible when the receiver sensitivity is very low (i.e., around -90~dBm) and the spacing between nanonodes in increased to relatively high values.
This is depicted in Figure~\ref{fig:rx_sensitivity}b) for the spacing of 3~mm. 
In such a case, the accuracy of the outliers is reduced, while the availability of the localization service experiences a rapid drop. 

%% file: conclusion.tex

\vspace{-1mm}
\section{Conclusion}

We have shown that the two-way \acf{ToF}-based trilateration has a potential for enabling accurate localization in \acf{THz}-operating energy harvesting \acfp{SDM}. 
We base our indication on the sub-millimeter accuracy and high availability of localization for the system parameterizations expected in real-life SDM implementations. 
Moreover, we have qualitatively characterized the effects of several relevant system parameters.
Example-wise, we have shown that the utilized bandwidth significantly affects the localization accuracy, while the energy harvesting rate and location update period play an important role in the service availability.
The derived numerical characterizations have to be taken only as preliminary ``rule-of-thumb'' indications due to several simplifications made in this work.

Future work will be focused on deriving more accurate characterizations of the localization capabilities.
Specifically, we will consider other relevant SDM mobility patterns, such as the ones expected in smart textiles.
We will also aim at a more accurate modeling of the THz nanoscale channel by for example considering scattering loss from aerosols and surfaces on which SDMs are mounted.   
Moreover, we will consider new metrics such as the localization latency, as well as characterize the effects of other relevant system parameters including the number of controllers, number of TS-OOK pulses used for the two-way ToF estimation, and different types of energy harvesting sources. 
In addition, we will consider a more accurate energy consumption modeling (e.g., accounting for the energy consumed in idling), as well as different energy consumption patterns (e.g., data transmission in the operational phase).
We will also aim at providing a system-level solution for localization in SDMs by specifying the signaling requirements. 
For example, we assumed that two-way ToF measurements can be derived for each controller-nanonode pair.
In reality, deriving such measurements will require a protocol that defines a sequence according to the nanonodes should transmit signals for the ToF estimation.    

Despite the simplifications and constraints, we believe this work provides a new argument for the utilization of THz-based wireless solutions for the control and programming of SDMs, in contrast to the utilization of wired solutions.
This is because a THz-based solution, in addition to being form-factor and energy-wise superior, would allow localizing the metamaterial elements under mobility.  